\documentclass[prl,twocolumn,superscriptaddress,showpacs,showkeys,floatfix]{revtex4-1}

\usepackage{graphicx}
\usepackage{bm}
\usepackage{amsmath}

\begin{document}

\title{Structure, magnetic order and excitations in the 245 family of Fe-based superconductors}

\author{Wei Bao}
\email{wbao@ruc.edu.cn}
\affiliation{Department of Physics, Renmin University of China, Beijing 100872, China}

\begin{abstract}
Elastic neutron scattering simultaneously probes both the crystal structure and magnetic order in a material. 
Inelastic neutron scattering measures phonons and magnetic
excitations. Here we review the average composition, crystal structure and magnetic order in the 245 family of Fe-based superconductors and in related insulating compounds from neutron diffraction works. A three-dimensional phase-diagram summarizes various structural, magnetic and electronic properties as a function of the sample composition. High pressure phase diagram for the superconductor is also provided. Magnetic excitations and the theoretic Heisenberg Hamiltonian are 
provided for the superconductor. Issues for future works are discussed.
\end{abstract}


\maketitle

Most Fe-based superconductors are pnictides \cite{rev2009H}. There had been only one
family of iron chalcogenide superconductors Fe$_{1+\delta}$(Se,Te) of maximum $T_C\approx 14$ K at ambient pressure \cite{A072369,A074775} and 37 K at high pressure \cite{B032143},
before a new iron chalcogenide superconductor of the {\em nominal composition} K$_{0.8}$Fe$_2$Se$_2$ was reported in 2010 with $T_C\approx 30$ K \cite{C122924}. A transient transition around 40 K from a part of the sample was also reported in the work, which may be related to the
superconducting transition realized later in $A_x$Fe$_2$Se$_2$(NH$_2$)$_y$(NH$_3$)$_z$ ($A$=K, Li) \cite{F027178,E035046}.

The Fe$_{1+\delta}$Se (11) superconductor is made of charge neutral FeSe layers of the anti-PbO structure, with the excess Fe important to the stability of the structure \cite{A092058,A111613}. The weakly coupled layers is susceptible to intercalation.
We will review magnetic order and excitations of the 30 K intercalated chalcogenide superconductors from neutron scattering studies.
To do that, it is also necessary to review sample composition, crystal structure and phase diagram of the new family of Fe-based superconductors, which have caused much confusion and controversy at the moment due to inadequate sample characterization. Physical parameters of the five $A_{2}$Fe$_{4}$Se$_5$ (245) superconductors are summarized in TABLE~\ref{tab1}.

\begin{table}[b!]
\caption{Physics properties of the $A_{2}$Fe$_{4}$Se$_5$ superconductors.
Lattice parameters at 295 K, magnetic moment of Fe at 10 K ($M$) as well as the superconducting transition temperature $T_C$, antiferromagnetic transition temperature $T_N$ and Fe vacancy ordering temperature $T_S$ are quoted from \cite{D020830,D022882}.
}
\label{tab1}
\begin{tabular}{c|ccccc}
\hline \hline
 $A$ & K & Rb  & Cs  & Tl,K & Tl,Rb \\
    \hline
  a ($\AA$) & 8.7306(1) & 8.788(5) & 8.865(5) & 8.645(6) & 8.683(5)  \\
 c ($\AA$) & 14.1128(4)& 14.597(2) & 15.289(3) & 14.061(3)& 14.388(5)  \\
  $ M$ ($\mu_B$)& 3.31(2)  & 3.3(1)  & 3.3(1)& 3.2(1)  & 3.2(1)   \\
 $T_C$ (K) & 32 & 32 & 29 & 28 & 32 \\
 $T_N$ (K) & 559(2) & 502(2) & 471(4) & 506(1)  &  511(1) \\
$T_S$ (K) & 578(2) & 515(2) & 500(1) & 533(2)  & 512(4) \\    
\hline \hline
    \end{tabular}
\end{table}

\section{Sample composition and crystal structure of the superconductors}

After the initial report \cite{C122924} and confirmation \cite{C124950} of superconductivity in the nominal K$_{x}$Fe$_2$Se$_2$, Cs and Rb compounds of the similar nominal composition
\cite{C123637,C125525} as well as Tl containing compounds of a different nominal composition (Tl,$A$)Fe$_{2-x}$Se$_2$ ($A$=K,Rb) \cite{C125236,D010462} were reported to superconduct also at $T_C\approx 30$ K. The former composition formula indicates an intact FeSe plane as in the 11 compounds as well as ``heavy electron doping'' \cite{C125980}. The latter suggests Fe vacancy in the FeSe plane since two different kinds of Fe vacancy orders have been reported previously in chalcogenide TlFe$_{2-x}$S$_2$ and TlFe$_{2-x}$P$_2$ \cite{jmmm86,jssc86}.

The nominal K$_{0.8}$Fe$_2$Se$_2$ and Cs$_{0.8}$Fe$_2$Se$_2$ superconducting samples used in the ARPES study to conclude a heavy electron doping \cite{C125980}, however,
are refined to be K$_{0.775(4)}$Fe$_{1.613(1)}$Se$_2$ and Cs$_{0.748(2)}$Fe$_{1.626(1)}$Se$_2$ in the 
single-crystal x-ray diffraction study using the samples from the same source \cite{D014882}. One superconducting sample from a systematic synthesis
study of another group \cite{D010789} is found to be K$_{0.737(6)}$Fe$_{1.631(3)}$Se$_2$ in the single-crystal x-ray diffraction study \cite{D014882}, while another superconducting sample to be K$_{0.83(2)}$Fe$_{1.64(1)}$Se$_2$ in a powder neutron diffraction refinement study \cite{D020830}.
Thus, the iron valance in the superconducting samples is very close to but not exactly at 2+, similar to that in the previously discovered iron chalcogenide superconductors Fe$_{1+\delta}$(Se,Te) \cite{A092058,A111613}, but very different from the conclusion of the ARPES studies which claim intact FeSe layers in $A_{0.8}$Fe$_2$Se$_2$ \cite{C125980,D014556,D014923}. An alternative interpretation of the ARPES data has been offered by Berlijn et al~\cite{E042849}.

\begin{figure}[tbh!]
\includegraphics[width=.65\columnwidth]{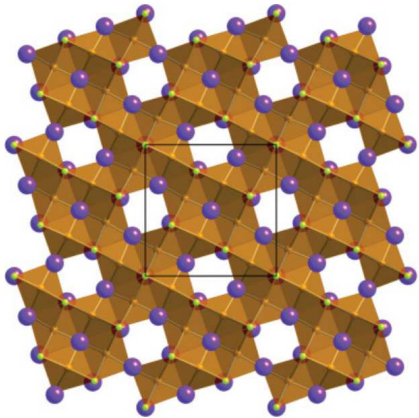}
\caption{Crystal structure of $A_{2}$Fe$_{4}$Se$_5$ from the [001] direction in the $I4/m$ unit cell showing fully occupied Fe2
sites (orange) decorated with ordered vacancy Fe1 sites. The $A$ site is represented by the blue ball, and the Se site by the yellow ball. (From Figure 3 of \cite{D014882})
} 
\label{vacancy_order1} 
\end{figure}

The Fe vacancy not only exists in these new superconductors but also forms a nearly ideal $\sqrt{5}$$\times$$\sqrt{5}$ superlattice on the FeSe square plane, see FIG.~\ref{vacancy_order1}.
The ordered structure is refined in the tetragonal $I4/m$ unit cell, and the structure parameters are tabulated in \cite{D014882,D020830}. The prominent structural feature of the superconducting samples is the {\em almost} empty 4$d$ Fe1 site with an occupancy $n(4d)$ at a few percent and the full occupancy of the 16$i$ Fe2 site, see TABLE~\ref{tab2}, below room temperature.

\begin{table*}[bth!]
\caption{Occupancy at the minority Fe1 site $4d$ and majority Fe2 site $16i$ at various temperatures for superconducting sample K$_{0.83}$Fe$_{1.64}$Se$_2$ and insulating sample K$_{0.93}$Fe$_{1.52}$Se$_2$, K$_{0.862}$Fe$_{1.563}$Se$_2$ and K$_{0.99}$Fe$_{1.48}$Se$_2$. The quantity $1-n(4d)/n(16i)$ can serve as a measure for the perfectness of the $\sqrt{5}$$\times$$\sqrt{5}$ vacancy order.
}
\label{tab2}
\begin{tabular}{c|cccc|c|c|c}
\hline \hline
 & \multicolumn{4}{|c|}{K$_{0.83(2)}$Fe$_{1.64(1)}$Se$_2$ \cite{D020830} }  & K$_{0.862(3)}$Fe$_{1.563(4)}$Se$_2$ \cite{D020488} & K$_{0.93(1)}$Fe$_{1.52(2)}$Se$_2$ \cite{D020488} & K$_{0.99(1)}$Fe$_{1.48(1)}$Se$_2$ \cite{D023674} \\
  T(K) & 11 & 295  & 500  & 550 & 90 & 100 & 295 \\
    \hline
  $n(4d)$ & 0.062(8) & 0.059(6) & 0.22(2) & 0.26(1) & 0.227(3) & 0.118(7) & 0.29(1)  \\
 $n(16i)$ & 1.008(4) & 1.020(6) & 0.951(4) & 0.935(3) & 0.920(1) & 0.918(5) & 0.857(6) \\
\hline
$n(4d)/n(16i)$ & 0.062(8) & 0.058(6) &0.23(2)&0.28(1)&0.247(3)&0.129(8)&0.34(1)\\
\hline \hline
    \end{tabular}
\end{table*}

If the Fe1 site is completely empty and the Fe2 site fully occupied with the perfect $\sqrt{5}$$\times$$\sqrt{5}$ Fe vacancy order, the sample composition would be $A_{0.8}$Fe$_{1.6}$Se$_2$, or $A_{2}$Fe$_{4}$Se$_5$ (245). As the non-stoichiometric Fe$_{1+\delta}$(Se,Te) (11) superconductors are referred to as the 11 family, the slightly off-stoichiometric iron selenide superconductors with the Fe vacancy order $A_{2+\epsilon}$Fe$_{4+\delta}$Se$_5$,
where $4\delta=n(4d)$ and $\epsilon\sim -2\delta$ due to the Fe valence $\sim 2$+, can also be referred to as the 245 superconductors \cite{D014882,D020830}. 

In addition to K and Cs, the $A =$ Rb, (Tl,K) and (Tl,Rb) superconducting $A_{2}$Fe$_{4}$Se$_5$ samples also possess the same nearly ideal $\sqrt{5}$$\times$$\sqrt{5}$ Fe vacancy order \cite{D022882}. The Fe vacancy order in all of these five known 245 superconductors disappears in an order-disorder structural transition at a very high temperature $T_S$ ranging from 500 to 578 K, respectively \cite{D020830,D022882}. 

Above the transition at $T_S$, the Fe1 and Fe2 sites become equally occupied, thus restoring the $I4/mmm$ symmetry as in the BaFe$_2$As$_2$ (122) system. However, it should be noted that this high-temperature compound cannot be called 
$A$Fe$_2$Se$_2$ as many people mistakenly do in current literature, since the occupancy at the Fe and $A$ sites in the $I4/mmm$ structure is only around 0.8 in these superconductors, see Table 2 in \cite{D020830}. The sample composition remains to be close to $A_{0.8}$Fe$_{1.6}$Se$_2$, i.e. 245. In other words, 245 does not always have the $\sqrt{5}$$\times$$\sqrt{5}$ superlattice structure. Temperature is one determining factor, more on this point in the phase-diagram section. A 245 compound at high temperature above $T_S$ does not make it a 122 compound, despite of its share of the same space group symmetry $I4/mmm$ with the 122 compounds.

\section{Magnetic order of the 245 superconductors}

\begin{figure}[tb!]
\includegraphics[width=\columnwidth]{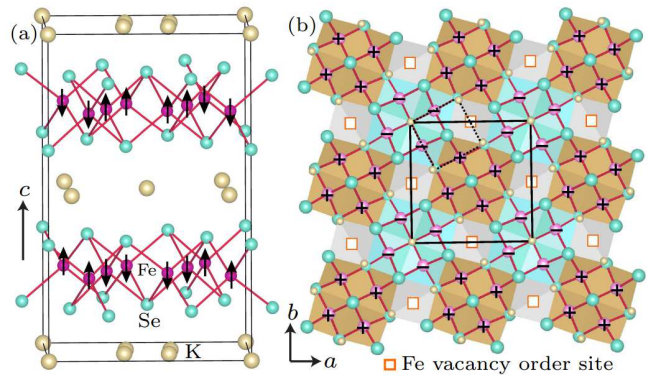}
\vskip -.3cm
\caption{(a) Magnetic structure of K$_2$Fe$_4$Se$_5$ in the $I4/m$ unit cell. 
(b) The in-plane antiferromagnetic ordering pattern of the four-spin blocks is highlighted with the two different colors that represent the alternating moment directions along the $c$-axis. 
(From Figure 3 of \cite{D020830})
} 
\label{AF_order1} 
\end{figure}

Slightly below the order-disorder transition $T_S=578$ K, an antiferromagnetic order develops at $T_N=559$ K in the K$_{2}$Fe$_{4}$Se$_5$ superconductor \cite{D020830}. The c-axis is the easy axis of the magnetic order. The four magnetic moments on the nearest-neighbor square block behave like a superspin, forming the simple chessboard antiferromagnetic pattern in the plane, see FIG.~\ref{AF_order1}. With the development of the staggered magnetic moment which reaches 3.31(2) $\mu_B$/Fe at 11 K, the distance between the four Fe atoms on the block of the same spin orientation also shrinks from the regular square lattice in the high temperature $I4/mmm$ structure \cite{D020830}. This strong
magnetostructural tetramerization greatly contributes to the stability of the block antiferromagnetic order on the $\sqrt{5}$$\times$$\sqrt{5}$ vacancy ordered lattice according to band structure calculations \cite{D021344,D022215}.

\begin{figure}[bt!]
\includegraphics[width=.85\columnwidth]{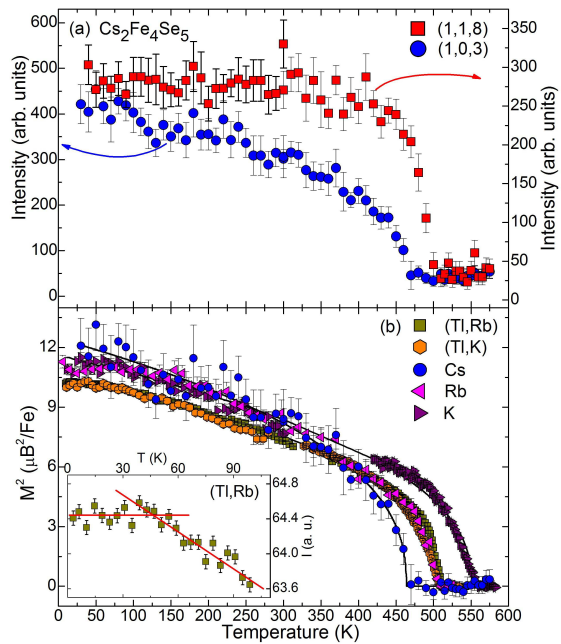}
\vskip -.3cm
\caption{(a) Magnetic (103) and structural (118) Bragg peaks vs the temperature, serving as order parameters for
the antiferromagnetic and Fe vacancy order-disorder transitions, respectively, in Cs$_2$Fe$_4$Se$_5$. 
(b) Normalized magnetic Bragg intensity representing the squared magnetic moment as a function of the temperature for the five 245 superconductors. 
Inset: Magnetic (101) peak of (Tl,Rb)$_2$Fe$_4$Se$_5$. The intensity saturates when $T_C$ is approached.
(From Figure 3 of \cite{D022882})
} 
\label{order_para} 
\end{figure}

The same large moment and high $T_N$ antiferromagnetic order in FIG.~\ref{AF_order1} also exists in the rest four 245 superconductors \cite{D022882}. In FIG.~\ref{order_para}(a), magnetic Bragg peak (103) due to the tetramer block antiferromagnetic order is shown as a function of temperature together with the structural Bragg peak (118) due to the $\sqrt{5}$$\times$$\sqrt{5}$ Fe vacancy order. The magnetic order starts to develop as soon as sufficient order has been established in the $\sqrt{5}$$\times$$\sqrt{5}$ superlattice. 
FIG.~\ref{order_para}(b) compares the squared magnetic order parameter of the five 245 superconductors. The $T_N$ ranges from 471 to 559 K but the staggered magnetic moment stays at 3.3(1) $\mu_B$/Fe, close to the atomic value 4 $\mu_B$/Fe for the Fe$^{2+}$ ions, refer to TABLE~\ref{tab1}. The staggered magnetic moment in 245 superconductors is larger than the record size 2.0 $\mu_B$/Fe of previous Fe-based superconductors \cite{A092058}.

One remarkable feature of the 245 superconductors is the coexistence of superconductivity with the very strong antiferromagnetic order.
Inset to FIG.~\ref{order_para} shows the inflection of the magnetic intensity when $T_C$ is approached in (Tl,Rb)$_2$Fe$_4$Se$_5$ superconductor. This indicates strong interaction between the antiferromagnetic order and superconductivity, and has served as a definite evidence for the coexistence in unconventional magnetic superconductors such as heavy-fermion UPt$_3$ \cite{UPt3_GA} and Fe-based
Ba(Fe$_{0.953}$Co$_{0.047}$)$_2$As$_2$ \cite{B032833}. The superconducting symmetry of the 245 materials has to be such that it can survive in the tremendously strong staggered magnetic field imposed by the large magnetic moment.

While the antiferromagnetic order in 245 superconductors does not break the four-fold tetragonal crystal symmetry, antiferromagnetic order in all previous families of Fe-based superconductors exists in a distorted crystal structure of a symmetry lower than the tetragonal one \cite{A040795,A062776,A092058}. It has been discovered in neutron diffraction works that there exists {\em empirical rules} that defines the relation between the shortened spacing of the Fe neighboring pair and their ferromagnetic interaction, and relation between the expanded spacing and antiferromagnetic interaction in NdFeAsO (1111) \cite{A062195}, BaFe$_2$As$_2$ \cite{A062776} and Fe$_{1+\delta}$Te \cite{A092058} of the 1111, 122 and 11 families. This intimate structure-magnetism relationship can be explained as due to different occupation of the $d_{xz}$ and $d_{yz}$ orbitals, which leads to the structural distortion from the in-plane four-fold symmetry and prepares the magnetic exchange interactions of correct signs for the observed antiferromagnetic order \cite{A062195,A042252,A062776,A092058}. Such an orbital ordering mechanism can also be successfully applied to explain the tetramer block antiferromagnetism in 245 superconductors in a unified fashion \cite{D050432,E060881}, satisfying the same empirical rules connecting the lattice expansion (contraction) with (anti)ferromagnetic exchange interaction \cite{D020830}. Meanwhile, the spin-density-wave scenario due to the nesting Fermi surface \cite{A040795} faces serious experimental difficulties, such as no anomaly in resistivity at $T_N$. Refer to \cite{F087405} for more detailed discussion.

\begin{figure}[t!]
\includegraphics[width=.75\columnwidth]{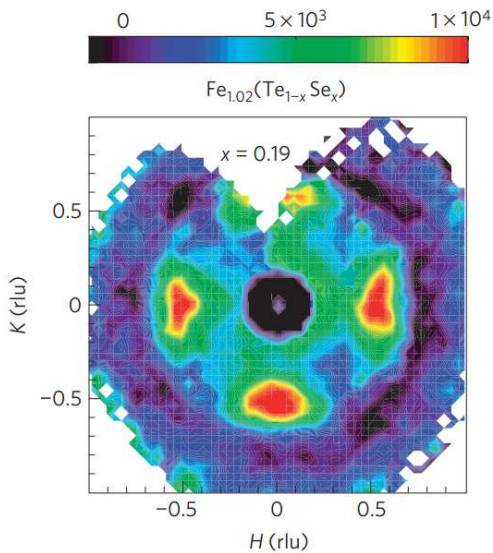}
\vskip -.3cm
\caption{Neutron scattering intensity map
on the (HK0) plane for Fe$_{1+\delta}$(Se,Te) in the tetragonal phase \cite{C035647}.
Magnetic clusters containing the four-spin block as that in 245 iron chalcogenides are responsible for the unusual intensity distribution, refer to \cite{D095196} for details.
(From Figure 4a of \cite{C035647})
} 
\label{diffuse} 
\end{figure}

When the lattice distortion of Fe$_{1+\delta}$Te is suppressed with Se substitution,
the long-range antiferromagnetic order that breaks the four-fold symmetry is replaced by a glassy short-range magnetic order \cite{A092058}. FIG.~\ref{diffuse} shows the diffuse scattering pattern in the basal plane from the glassy order. The order contains the same tetramer block of 245 superconductors as a constituent in fluctuating magnetic clusters \cite{D095196,D035073}. Thus, it appears that the magnetostructural tetramerization is a common tendency among the two families of iron chalcogenide superconductors \cite{E060881}. The condensation of the tendency to a long-range order possibly needs the relief of magnetic frustration that is alluded to by Yildirim \cite{A042252}, with the lattice distortion in the 11 family or the tetramerization introduced by the $\sqrt{5}$$\times$$\sqrt{5}$ vacancy order in the 245 family.

\section{Composition Phase Diagram} 

No polycrystalline 245 superconductor has been reported so far. The as-grown single crystal is also usually 
non-superconducting.
The crystal becomes superconducting after an annealing process. When the superconducting sample is ground, we always find pure Fe intergrowth between the 245 plates \cite{D020830}. This may not be surprising for intercalating compounds. However, it has contributed to widespread systematic error in the determination of the sample composition in current research, that derives the nominal composition either 
using the starting material ratio, or employing the inductively coupled plasma (ICP) analysis method when both the genuine 245 crystal and the Fe intergrowth are vaporized. 
In neutron or x-ray diffraction experiments, different materials in the bulk sample can be detected, and their compositions and structures separately refined. We have found that the refined composition is quite different from the nominal composition \cite{D014882,D020830}.

\begin{figure}[tb!]
\includegraphics[width=\columnwidth]{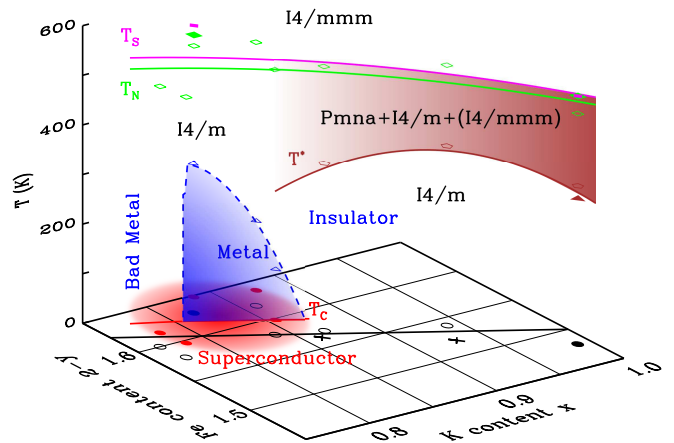}
\caption{The phase diagram of K$_{x}$Fe$_{2-y}$Se$_2$. 
In the $I4/mmm$ phase, Fe vacancy is randomly distributed.
In the order-disorder transition at $T_S$, Fe vacancy forms the $\sqrt{5}$$\times$$\sqrt{5}$ superlattice of space group $I4/m$ for samples in the neighborhood of K$_{0.8}$Fe$_{1.6}$Se$_2$ on the left, 
and forms the orthorhombic superlattice of space group $Pmna$ [FIG.~\ref{patterns}(c)] and imperfect $\sqrt{5}$$\times$$\sqrt{5}$ superlattice of space group $I4/m$ together with remnant disordered $I4/mmm$ phase on the right in the shaded phase-separation region.
Below $T^*$, the $Pmna$ and $I4/mmm$ phases transform to the imperfect $\sqrt{5}$$\times$$\sqrt{5}$ Fe vacancy order.
The occupancy ratio of the Fe1/Fe2 sites measures the perfectness of the $\sqrt{5}$$\times$$\sqrt{5}$ Fe vacancy order. When the ratio approaches zero, insulator-metal crossover occurs in the blue region,
beneath which superconductivity occurs at $T_C$ (red symbols). $T_N$ (green symbols) marks the antiferromagnetic transition.
(From Figure 4 of \cite{D023674})
} 
\label{phase_D} 
\end{figure}

FIG.~\ref{phase_D} shows the phase diagram of K$_{x}$Fe$_{2-y}$Se$_2$, of which samples were prepared
using the same procedure for the series of samples published in \cite{D010789}. First of all, the $x$ and $y$ in the chemical formula K$_{x}$Fe$_{2-y}$Se$_2$ are not completely independent, and the actual sample compositions shown on the basal plane cluster along the black line $x=2y$ which defines the Fe valance exactly at $2+$. Thus, the series of samples made using the Bridgman method all contain Fe of valence close to $2+$. Consequently with $x\approx 2y$, the charge neutrality requires that when more K of valence $1+$ is intercalated between the FeSe planes, more Fe vacancy is created. 
When $x=0.8$ and $y=0.4$ ($2-y=1.6$) at the left side of the phase-diagram, K$_{x}$Fe$_{2-y}$Se$_2$ becomes K$_{0.8}$Fe$_{1.6}$Se$_2$, namely K$_{2}$Fe$_{4}$Se$_5$. One of five Fe sites is vacant. 
When $x=1$ and $y=0.5$ at the right, K$_{x}$Fe$_{2-y}$Se$_2$ is KFe$_{1.5}$Se$_2$, i.e. K$_{2}$Fe$_{3}$Se$_4$ (234). One of four Fe sites is vacant. 

\begin{figure*}[bt!]
\includegraphics[width=1.5\columnwidth]{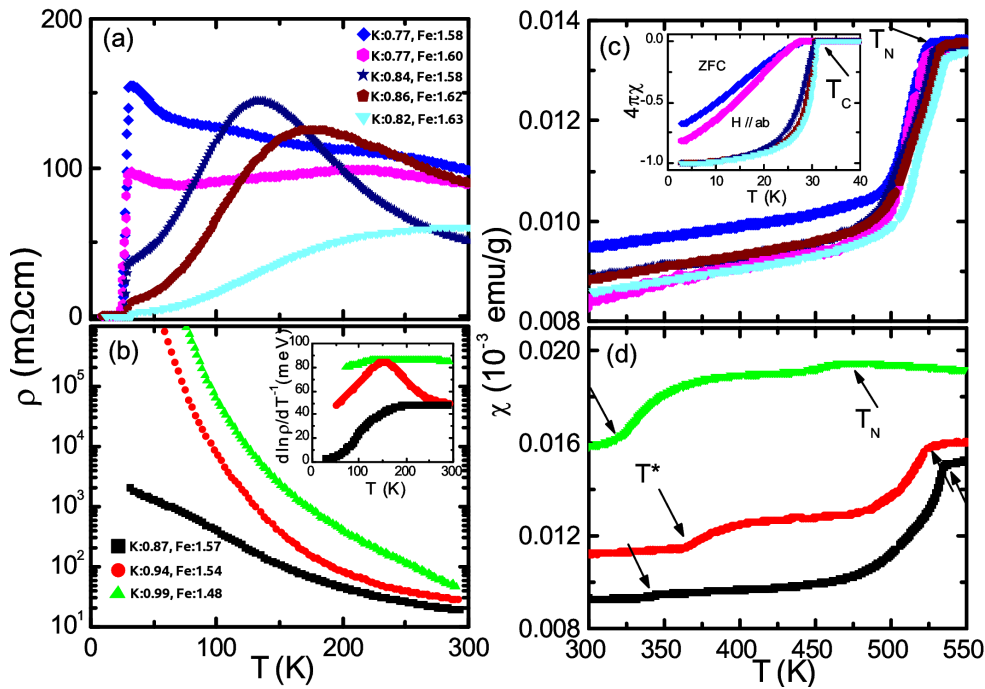}
\caption{The resistivity of (a) superconducting and (b)
insulating K$_{x}$Fe$_{2-y}$Se$_2$ samples. The inset to (b) is the activation gap in transport, which closes at low temperature when $x$ is reduced to
0.86. The magnetic susceptibility of (c) the superconducting and (d) the insulating samples, showing
antiferromagnetic transition at $T_N$ and the disappearance of the one-out-of-four orthorhombic Fe vacancy order at $T^*$.
(From Figure 3 of \cite{D023674})
} 
\label{bulk} 
\end{figure*}

The two superconducting samples at left in the phase diagram FIG.~\ref{phase_D} show consistently high a resistivity above $T_C$, see FIG.~\ref{bulk}(a). They also show poor diamagnetic response below $T_C$, see inset to FIG.~\ref{bulk}(c). Moving to the right in the phase diagram, the normal state resistivity of the next three superconducting sample shows a bump, which defines the metal-insulator crossover point (blue triangle and dashed line) in the phase diagram. These samples demonstrate a much better diamagnetic response. The sample K$_{0.83}$Fe$_{1.64}$Se$_2$ of the highest crossover temperature, showing the lowest overall normal state resistivity, was made following the same recipe as that used in the previous neutron powder diffraction study \cite{D020830}. Below the crossover temperature where the metal-like positive-slope resistivity exists, the occupancy ratio of the two Fe sites $n(4d)/n(16i)\approx 0.06$ has approached the minimum value that we have observed, see TABLE~\ref{tab2}, indicating a highly ordered $\sqrt{5}$$\times$$\sqrt{5}$ vacancy superlattice.

Moving further to the right in the phase diagram FIG.~\ref{phase_D}, the three insulating samples shows a progressive opening of a transport activation gap from a logarithmic behavior at low temperature, see FIG.~\ref{bulk}(b) and inset therein. Except under the small dome of the metallic crossover and the superconducting phase underneath, most region of the phase diagram stays in insulating phase \cite{D023674}.

Basca et al.\ are the first to perform single crystal x-ray refinement study of K$_{0.862(3)}$Fe$_{1.563(4)}$Se$_2$ at 90 K and K$_{0.93(1)}$Fe$_{1.52(2)}$Se$_2$ at 100 K \cite{D020488} in this insulating phase, and their compositions are marked by the crosses in the basal plane of FIG.~\ref{phase_D}.
The crystal structure in the temperature range is described by the $I4/m$ space group. However, there is substantial disorder in the $\sqrt{5}$$\times$$\sqrt{5}$ superstructure. The refined structure parameters for the first crystal are list in Table 1 in \cite{D020488}. The occupancy at the two Fe sites is quoted in TABLE~\ref{tab2}.
The $n(4d)/n(16i)=0.247(3)$ for the first crystal and $0.129(8)$ for the second crystal.

\begin{figure}[b!]
\includegraphics[width=\columnwidth]{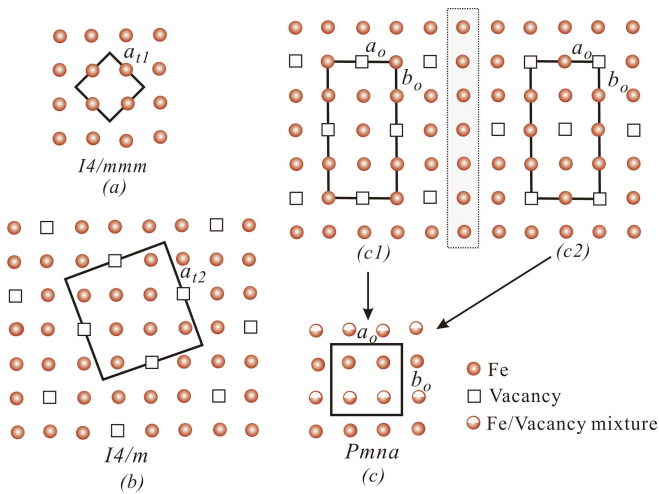}
\caption{An Fe layer in the (a) $I4/mmm$,
(b) $I4/m$ and (c) $Pmna$ structure, respectively, with the
solid line marking the unit cell. With the perfect order shown in (a), (b) or (c1), the sample composition would be
KFe$_2$Se$_2$ , K$_2$Fe$_4$Se$_5$ or K$_2$Fe$_3$Se$_4$, respectively. The
vacancy order in (c) is an average of those in (c1) and (c2).
(From Figure 2 of \cite{D023674})
} 
\label{patterns} 
\end{figure}

We performed neutron powder diffraction study in a wide temperature range for K$_{0.99(1)}$Fe$_{1.48(1)}$Se$_2$, which is very close to the 234 end member \cite{D023674}. One of four Fe is vacant at the 234 composition, and the vacancy order shown in FIG.~\ref{patterns}(c1) or (c2) has been discussed in works on TlFe$_{1.5}$Se$_2$ and TlFe$_{1.5}$S$_2$ \cite{jmmm86}. Surprisingly, this orthorhombic
superstructure is not the ground state in KFe$_{1.5}$Se$_2$. 
It appears in the order-disorder transition at $T_S\approx 500$ K down to a finite temperature $T^*\approx 295$ K as a competing phase coexisting with an imperfect $\sqrt{5}$$\times$$\sqrt{5}$ Fe vacancy order, as well as with a remnant vacancy-disordered phase of the $I4/mmm$ symmetry from high temperature \cite{D023674}. As one of the three phases in the phase separated region in the phase diagram FIG.~\ref{phase_D}, substantial structural faults exist in this vacancy order so that the average pattern in FIG.~\ref{patterns}(c) describes our data. 
The refined
structural parameters in orthorhombic $Pmna$ space group are listed in Table 3 in \cite{D023674}.
Magnetic structure in the orthorhombic phase is the same as that in the BaFe$_2$As$_2$ with the staggered moment
2.8(1) $\mu_B$/Fe from neutron diffraction study on K$_{0.85}$Fe$_{1.54}$Se$_2$ \cite{E041316}.

Below 380 K, no remnant $I4/mmm$ phase can be detected in K$_{0.99}$Fe$_{1.48}$Se$_2$.
Below $T^*\approx 295$ K, the $Pmna$ phase also disappears, leaving the $\sqrt{5}$$\times$$\sqrt{5}$
Fe vacancy order as the only phase at low temperature \cite{D023674}, consistent with the case of 
K$_{0.93}$Fe$_{1.52}$Se$_2$ at 100 K and K$_{0.862}$Fe$_{1.563}$Se$_2$ at 90 K in the x-ray work \cite{D020488}. Refined structure parameters for K$_{0.99}$Fe$_{1.48}$Se$_2$ at 50 and 295 K are listed in
Table 2 of \cite{D023674}, and the staggered antiferromagnetic moment is 3.16(5) $\mu_B$/F, similar to the value in 245 \cite{D020830}. The Fe occupancy data for K$_{0.99}$Fe$_{1.48}$Se$_2$ are also quoted in TABLE~\ref{tab2}. Together with the data from Basca et al.\ \cite{D020488}, the enhanced $n(4d)/n(16i)$ in the insulating phase of FIG.~\ref{phase_D} reflects the increasing disorder necessary to resolve the mismatch between the number of Fe vacancies in the material and the number of vacancies in the $\sqrt{5}$$\times$$\sqrt{5}$ superlattice pattern.

\begin{figure*}[t!]
\includegraphics[scale=.94]{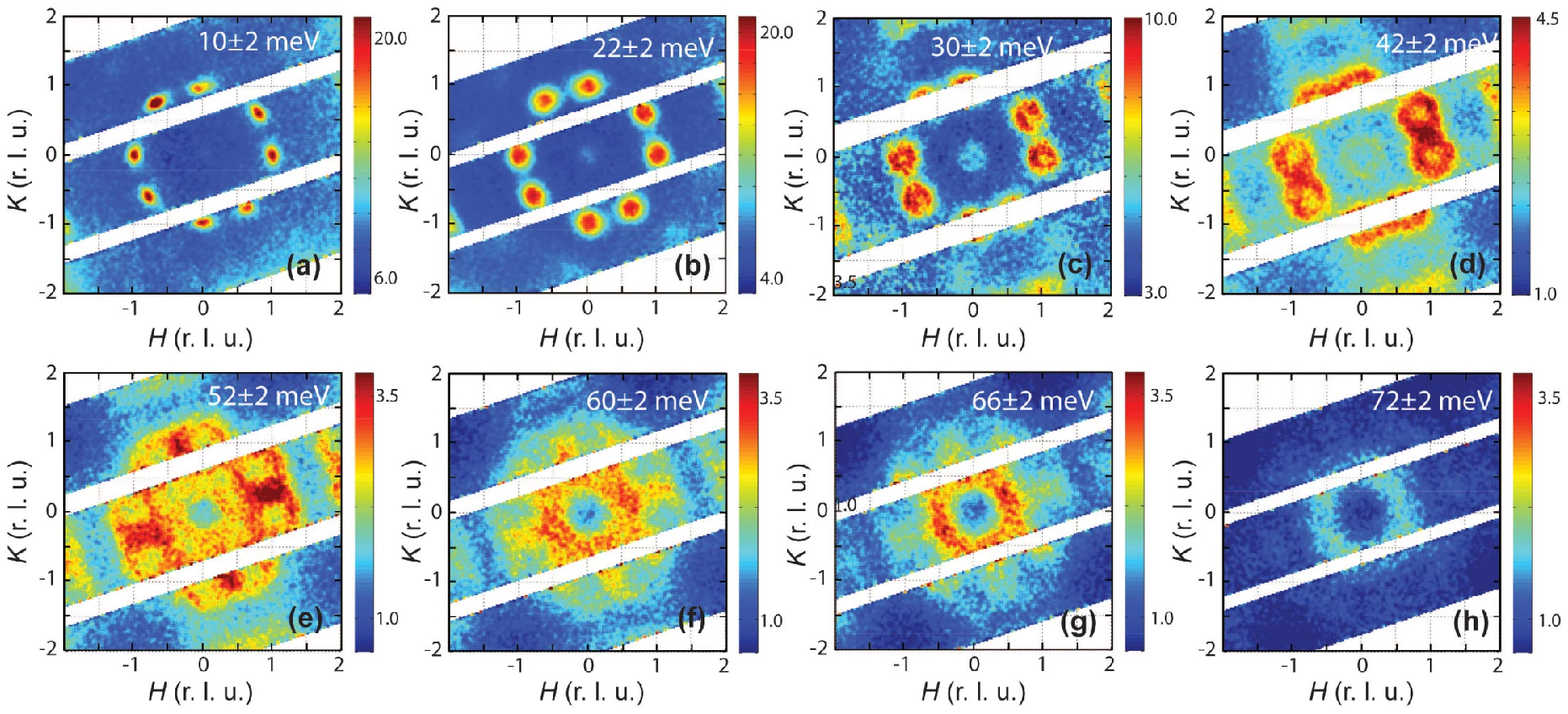}
\caption{Constant energy slices of the acoustic branch of the
spin wave excitations projected on the $(H,K,0)$ plane. The energy transfer is specified on each figure.
The relative intensity is indicated by the color scale. 
The sample was aligned on one of the two crystalline twins in the $I4/m$ unit cell.
The data were collected at SEQUOIA chopper spectrometer with $E_i = 50$ meV for (a-b) and 100 meV for
the other panels.  (From Figure 2 of \cite{E012413})
}
\label{fig8}
\end{figure*}

The site disorder registered in the substantial $n(4d)/n(16i)$ value for samples away from 245 on the right part of the phase diagram in FIG.~\ref{phase_D} manifests in the telltale logarithmic resistivity close to the metal-insulator crossover shown in FIG.~\ref{bulk}(b). Above the crossover temperature for the 245 sample, the $n(4d)/n(16i)$ value also increases substantially from the 0.06 base value, see TABLE~\ref{tab2}.
Therefore, the metal-insulator crossover in the K$_{x}$Fe$_{2-y}$Se$_2$ system is likely also driven by the
Anderson weak localization process. A similar situation in the 11 chalcogenide superconductors, albeit not by site disorder scattering but by spin-glass quasi-static scattering of electrons, has been uncovered
previously in a similar study combining neutron scattering and bulk techniques \cite{C035647}.
In both 11 and 245 iron chalcogenide superconductors, therefore, metallic transport behavior in the normal state serves as the prelude to superconductivity at lower temperature.

Neutron diffraction refinement studies not only provide microscopic mechanism for the normal state transport properties, they also explain the anomalies in magnetic susceptibility. The antiferromagnetic transition leaves a strong reduction in the susceptibility, FIG.~\ref{bulk}(c)-(d). The further reduction at $T^*$ in FIG.~\ref{bulk}(d) corresponds to a rapid increase in the staggered magnetic moment from 2.8 to 3.2 $\mu_B$ in the conversion of the $Pmna$ to the $I4/m$ phase, see Fig.~1(b) in \cite{D023674}. The inhibition of the susceptibility is caused by the large 6.5(3) meV magnetic excitation gap of the block antiferromagnetic order \cite{E012413}.

The preference of the tetragonal $\sqrt{5}$$\times$$\sqrt{5}$ superlattice of the $I4/m$ symmetry over the orthorhombic one-out-of-four superlattice of the $Pmna$ symmetry shown in FIG.~\ref{patterns} as the ground state even in the 234 compound is an experimental confirmation to
the large electronic energy gain in the formation of the magnetostructural tetramers in {\it ab initio} energy band theory \cite{D021344,D022215}. When temperature is raised above $T_S$, the material enters a two-phase mixture of the two Fe vacancy orders to gain in entropy, henceforth in free energy, as can be understood in a simple two-level statistical physical model. 

A phase diagram with samples scattering along a bending line on the basal plane of FIG.~\ref{phase_D} has also been published that is consistent with our result \cite{D044941}. However, only $T_S$, $T_N$ and $T_C$ are marked, using features in bulk measurement data at the structural and magnetic transitions identified in our neutron scattering works. The use of the valence as the x-axis in their two-dimensional phase diagram can also be misleading, since a valence change originating at e.g. 234 will not lead to superconductivity.

The phase separated region between $T_S$ and $T^*$ in phase diagram FIG.~\ref{phase_D} covers the room temperature.
Many studies on 245 superconductors reporting phase separation used samples of actual compositions in this miscibility gap. Therefore,
all of the three types of structural phases in the phase region have been observed in transmission electron microscopy study \cite{D012059}. The existence of the $\sqrt{5}$$\times$$\sqrt{5}$ Fe vacancy order also shows up in new phonon modes detected in optic and Raman measurements of various samples over the phase-diagram \cite{D010572,D012168}.

\section{Magnetic excitations of the 245 superconductors}

The whole magnetic excitation spectrum in the (Tl,Rb)$_2$Fe$_4$Se$_5$ superconductor
(T$_C \approx 32$ K) has been measured up to 300 meV with a chopper inelastic neutron spectrometer \cite{E012413}. Some of the data are shown in FIG.~\ref{fig8}, demonstrating the evolution of the spin-wave cones at the two sets of magnetic Bragg spots from a twinned single-crystal sample.

\begin{figure}[t!]
\includegraphics[width=.9\columnwidth]{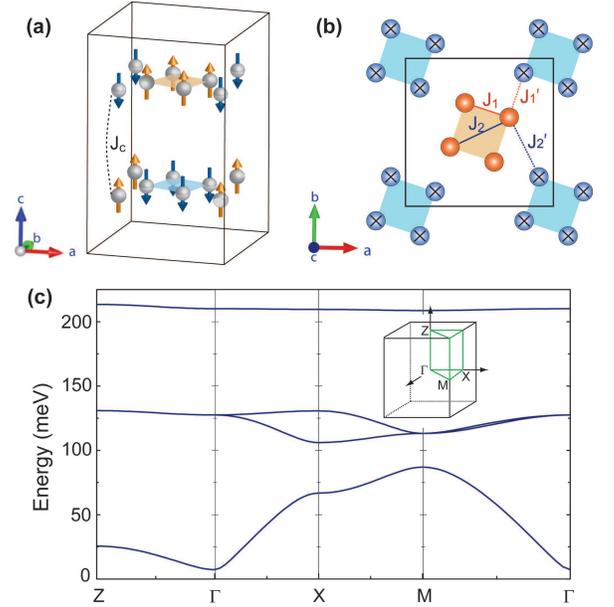}
\caption{(a)-(b) Schematic diagram showing 
$J_c$, the exchange interaction between 
spins in adjacent Fe planes, and the four unique in-plane
exchange interactions considered in this work. 
(c) Theoretical spin wave dispersions calculated using experimentally determined
parameters. (From Figure 1 of \cite{E012413})
} 
\label{Fig9} 
\end{figure}

Consistent with the expectation for a large-moment antiferromagnet, magnetic excitation spectrum
can be fitted by a Heisenberg Hamiltonian of localized magnetic moments
\begin{equation}
H =  \sum_{i,j} J_{i,j} S_{i} \cdot S_{j} - \Delta \sum_i S^2 _{iz},      
\label{Fig8}           
\end{equation}
which includes five exchange constants $J_1$, $J_2$, $J_1^{\prime}$, $J_2^{\prime}$ and $J_c$ 
as depicted in Fig.~\ref{Fig9}(a) and (b), and the single-ion anisotropy constant $\Delta$ that 
quantifies the observed Fe spin $S=3.2(1)/g$ alignment along the c-axis \cite{D022882}.
The spin-wave dispersion, as well as the scattering intensity of the acoustic branch, can be accounted for with the following parameters \cite{E012413}:
\begin{align}
SJ_1	&=	-30(1)  {\rm meV}, & SJ_1^{\prime} &= 31(13)  {\rm meV}, \nonumber \\ 
SJ_2&=10(2)   {\rm meV}, & SJ_2^{\prime} &= 29(6)  {\rm meV},\nonumber \\ 
SJ_c&=0.8(1)  {\rm meV}, & S\Delta &= 0.3(1)   {\rm meV}.
\end{align}
Except the weaker antiferromagnetic 
$J_2$ which frustrates the ferromagnetically aligned spin block, the remaining terms help stabilizing 
the observed block antiferromagnetic order. There exists a qualitative agreement between these experimental values and {\it ab initio} linear response theoretic results \cite{E056404}.
The resulting spin wave dispersion curves in various high symmetry directions are shown in
Fig.~\ref{Fig9}(c). 

Low energy magnetic excitations ($\lesssim 30$ meV) from a superconducting Rb$_{0.82}$Fe$_{1.68}$Se$_2$
sample has also been investigated with a chopper inelastic neutron spectrometer \cite{E013348}.
The data are not sufficient to determine the magnetic exchange interactions on the $\sqrt{5}$$\times$$\sqrt{5}$ Fe vacancy lattice, and the focus of the work was on some extra features around the in-plane ($\pi,0$) point in the $I4/mmm$ unit cell notation. No such features are observed in our study on the (Tl,Rb)$_2$Fe$_4$Se$_5$ superconductor. Judging from the phase-diagram in FIG.~\ref{phase_D} and the fact that the magnetic zone center of the $Pmna$ phase is at ($\pi,0$) \cite{E041316}, the Rb$_{0.82}$Fe$_{1.68}$Se$_2$ sample used in the work is likely not pure and the $Pmna$ impurity phase contributes the extra magnetic excitations.

The acoustic branch of magnetic excitations ($E \lesssim 80$ meV) from the K$_2$Fe$_4$Se$_5$ superconductor
has been measured with a triple-axis spectrometer along two directions in the reciprocal space \cite{F045950}.
An independent determination of magnetic Hamiltonian, however, is out of reach. Thus, the $J_1$ and $J_2$
used in data fitting were borrowed from the values of a neutron scattering study on an insulating Rb$_{0.98}$Fe$_{1.58}$Se$_2$ sample \cite{E054675}. This insulating sample locates at the right side of the phase diagram, see FIG.~\ref{phase_D}, thus necessarily containing substantial site disorder with a large $n(4d)/n(16i)$ value.
Although an exchange-interaction between the third nearest neighbor spin pair $J_3$ in the plane in principle is
possible \cite{E054675} in this insulating sample of rather disordered $\sqrt{5}$$\times$$\sqrt{5}$ superlattice, the argument to include $J_3$ in the intensity fitting could be nullified due to the inclusion of the scattering intensity from the close-by twin Bragg spot in the coarse spatial resolution volume of the spectrometer used in the study.
 
We do not review inelastic neutron scattering works on the so-called "resonance mode" in the 245 family of Fe-based superconductors. As shown in e.g.\ Figure 3 of \cite{F054898} for a related FeSe superconductor, the mode keeps decreasing above the superconducting transition $T_C$. This is fundamentally different from the resonance mode, observed with significant signal to noise ratio, in the 122 family of Fe-based superconductors. The association of the spectral feature with superconductivity is premature at the moment.

\section{Phase diagram at high pressure}

\begin{figure}[b!]
\includegraphics[width=.9\columnwidth]{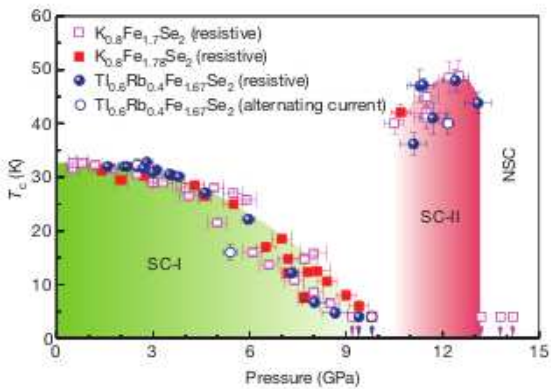}
\caption{Pressure dependence of $T_C$ for the 245 superconductors. (From Figure 4 of \cite{SunLL12})
} 
\label{sun_HP} 
\end{figure}

The stoichiometric $A_2$Fe$_4$Se$_5$ is an antiferromagnetic semiconductor according to band structure calculations \cite{D021344,D022215}. Superconducting samples, as presented above, are slightly 
off-stoichiometric and show metal-like transport property below the semiconductor-metal crossover temperature, FIG.~\ref{bulk}(a). The 245 superconductor also locates close to the miscibility gap of phase separation, refer to FIG.~\ref{phase_D}.
Thus, an important question is whether the 245 superconductor is a doped semiconductor or only a minority phase in the phase-separated sample is superconducting \cite{E091650}. A closely related question is whether inhomogeneity is intrinsic to a off-stoichiometric superconductor or a pure superconducting phase can be identified and hopefully isolated \cite{D070412,E084159,E025446,E030286,D012059,D083006,E031533,E031834}. The majority view at the moment is that the $I4/m$ phase with the $\sqrt{5}$$\times$$\sqrt{5}$ Fe superlattice and the large-moment antiferromagnetic order is irrelevant to the superconductivity. The A$_2$Fe$_3$Se$_4$, AFe$_2$Se$_2$ or A$_x$Fe$_2$Se$_2$
phase has all been proposed as the superconducting phase.

While the samples of the 245 family of superconductors involves complex preparation issues due to the proximity to the miscibility gap, high pressure offers another way to control the phases and investigate physics properties of the material \cite{D010092,D106138}. In particular, superconductivity
in (Tl,Rb)$_2$Fe$_4$Se$_5$ is suppressed by high pressure at $P_c \sim 9$ GPa, and then reappears between 11 and 13 GPa with a higher $T_C \approx 48$ K \cite{SunLL12}, FIG.~\ref{sun_HP}.

\begin{figure}[t!]
\includegraphics[width=.9\columnwidth]{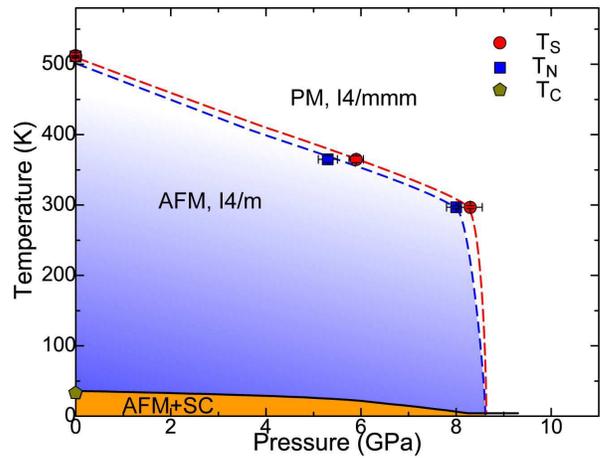}
\caption{Pressure-temperature phase diagram of $\rm (Tl,Rb)_2Fe_4Se_5$, showing the $\sqrt{5}$$\times$$\sqrt{5}$ superlattice transition at $T_S$, block antiferromagnetic transition at $T_N$ and superconducting transition at $T_C$. (From Figure 4 of \cite{G032624})} 
\label{HP245} 
\end{figure}

Single-crystal high pressure neutron diffraction study on the (Tl,Rb)$_2$Fe$_4$Se$_5$ superconductor has been recently performed \cite{G032624}. Both the $\sqrt{5}$$\times$$\sqrt{5}$ Fe vacancy order and the block antiferromagnetic order can be simultaneously measured. The phase diagram is shown in FIG.~\ref{HP245}. A note of caution is that neutron scattering experiment at 9 GPa is not possible to be conducted at very low temperature at this time.
A close relation of the structural and magnetic orders with the superconducting phase can be deduced.

High pressure x-ray structure study of the 245 superconductors have been conducted previously. The neutron diffraction result is consistent with that by Guo et al.\ \cite{D010092}, but fundamentally different from that by Ksenofontov et al.\ \cite{D123822}. Ksenofontov et al.\ also reported that the A$_x$Fe$_2$Se$_2$ phase survives beyond the pressure $P_c$ that suppresses the superconducting phase \cite{D123822}.

\section{Concluding remarks}

Elastic and inelastic neutron scattering studies have play a crucial role in determining the sample composition, crystal structure, phase diagram, magnetic order and excitations in the 245 family of Fe-based superconductors.
The well-ordered $\sqrt{5}$$\times$$\sqrt{5}$ superlattice characterized by a small $n(4d)/n(16i)\approx 0.06$ value in the $I4/m$ structure is linked
with the metal-like normal state transport property, which precedes the occurrence of the superconductivity. Like in the 11 family of iron chalcogenide superconductors, the Anderson localization of conducting electrons by disordered scattering is also fatal to superconductivity in the 245 family of iron chalcogenide superconductors.

The large-moment block antiferromagnetic order is crucial for the stability of the $\sqrt{5}$$\times$$\sqrt{5}$ Fe vacancy order over its competing phase such as
the orthorhombic vacancy order or the disordered Fe partial occupancy in the $I4/mmm$ structure, which exists only in an intermediate temperature range. This is due to the large energy gain in the tetramerization process. The same kind of magnetostructural coupling
in which the contraction and expansion of the lattice are linked with ferromagnetic and antiferromagnetic exchange interactions, respectively, is universal in all Fe based superconducting families, reflecting the common $d$ orbital ordering mechanism.

The low charge density left after the majority of $d$ electrons form the large magnetic moment close to the atomic limit on Fe$^{2+}$, the off-stoichiometric
composition of the superconducting phase, and the close-by miscibility gap of multiple phases have caused complex
material control and characterization issues. Whether the 245 superconductors exist as a pure phase like in e.g.\ the heavy fermion superconductors or in an inhomogeneous matrix like in e.g.\ the cuprate superconductors is a open question. The high pressure
works suggest a symbiotic relation of the $\sqrt{5}$$\times$$\sqrt{5}$ vacancy order and the block antiferromagnetic order with the superconductivity.

Nonetheless, we know that the 245 superconductors are of a spin-singlet state as shown in the NMR study \cite{D011017}. The superconductivity exists in close proximity with the strong antiferromagnetic order as demonstrated by its strong effect on magnetic order parameter. Conversely, the superconducting order parameter in the 245 materials should be able to withstand the strong staggered magnetic field. Additionally, the energy gap on the Fermi surface can be measured with ARPES. Further progress in pinning down microscopic material phase of the 245 superconductors at the correct average composition region on the phase diagram demands careful experimental study on well controlled and characterized samples.

\section{acknowledgments}
We thank Xianhu Chen, Minghu Fang, Genfu Chen, Qingzhen Huang, Yiming Qiu, M.A. Green, Feng Ye, Songxue Chi, A.T. Savici, G.E. Granroth, M.B. Stone, A.M. dos Santos, Qingming Zhang, Weiqiang Yu, Nanlin Wang, Jianqi Li, Lilian Sun, Donglai Feng, Wei Ku, Zhongyi Lu, Chao Cao and Jianhui Dai for collaboration and discussions in the investigation on the 245 Fe-based superconductors.  
The work at RUC was supported by National Basic Research Program of China 
(Grant Nos. 2012CB921700 and 2011CBA00112) and the National Natural Science
Foundation of China (Grant Nos. 11034012 and 11190024).


%

\end{document}